\documentclass{article}

\usepackage{graphicx}

\usepackage{natbib}

\newcommand{\be}{\begin{equation}}
\newcommand{\ee}{\end{equation}}

\newcommand{\ba}{\begin{eqnarray}}
\newcommand{\ea}{\end{eqnarray}}
\newcommand{\om}{\omega}
\newcommand{\Alfven}{Alfv\'{e}n }

\newcommand\etal{\textit{et al.\ }}
\newcommand\eg{\textit{e.g.\ }}

\newcommand{\Bf}{{magnetic field\,}}
\newcommand{\Bfs}{{magnetic fields\,}}

\newcommand{\ms}{magnetosphere\,}
\newcommand{\LC}{light cylinder}

\begin{document}

\title{On generation of Crab giant pulses}
\author{Maxim Lyutikov\\
Department of Physics, Purdue University, \\
 525 Northwestern Avenue,
West Lafayette, IN
47907-2036 }

\maketitle

\begin{abstract}
 We propose that Crab  giant pulses are generated     on closed magnetic  field lines  near   the light cylinder  via  anomalous cyclotron resonance on  the ordinary mode.  Waves are generated in a set of fine, unequally spaced, narrow emission bands at frequencies much lower than a local cyclotron frequency. Location of emission bands is fitted to  spectral   structures  seen by Eilek \etal (2006).
 
  To reproduce the data, the  required  density  of plasma in the giant pulses emission region is much higher, by a factor $\sim 3 \times 10^5$,  than the minimal Goldreich-Julian density. Emission is generated  by 
a population of highly energetic particles with radiation-limited Lorentz factors $\gamma \sim 7 \times 10^7$,  produced  during occasional reconnection close to the Y point, where  the  last closed field lines approach the  \LC.
\end{abstract}


\section{Introduction}

Giant pulses (GPs) are relatively rare durations 
of intense radio outbursts that are clearly a special form of 
pulsar radio emission \cite[\eg][]{Kuzmin}.  
Overall, the properties of GPs are (i) peak fluxes densities can exceed thousands (and,  in case  of Crab.  nearly a  million)
of times the peak flux
density of regular pulses (the Crab pulsar was discovered by observations of its giant pulses \citep{CrabDisc}); on average, energy coming in GPs can {\it exceed} the energy in an average pulse. Distribution of peak fluxes is power-law (unlike regular radio pulses which have 
Gaussian  distribution); (ii) GPs typically have short duration   from several microseconds down to few nanoseconds \citep{Hankins03}  (much shorter that sub-pulses in the average pulse) and emit narrowband radiation \citep{popov}; 
at lower time resolution, overlap of many narrowband emission spectra may mimic a broadband spectrum ; (iii) GPs seem to be limited to the edges of main profiles \citep[\eg][]{popov06} (see, though, 
unconfirmed observations of GP in HFCs by \cite{jessner}); (iv) GPs are seen in only 11 pulsars;  (v) there is  no correlation between GPs and emission at higher frequencies (nothing in  X-rays  \citep{Lund} and a marginal $3\%$  correlation in the optical  \citep{shearer03}); 
(vi) there seems to be  no strong  correlation between pulsar properties (period, \Bf, \Bf at the light cylinder, luminosity) and GP phenomenon.

In addition, recent observation of Crab pulsar \citep{eilek07} identified unique features of GPs associated with the interpulse (IGPs): (i) IGPs spectra consist of a number of relatively narrow frequency bands; (ii)  spacing between the bands is proportional, $\Delta \nu/\nu \sim 0.06$; (iii)  emission at different bands start nearly simultaneous, perhaps with a small delay at lower frequencies; (iv)  sometimes  there is a slight drift up in frequencies; all bands drift together, keeping the separation nearly constant; (v)  emission bands are located at $4-10$ GHz, continuing, perhaps, to higher frequencies, but {\it not}  to lower frequencies; (vi) all IGPs show band structure. These are very specific properties that allowed us  to build a {\it quantitative} model of pulsar radio emission (applicable, at least, for Crab GPs associated with interpulse).

Theoretically,    GPs  were proposed to be nonlinear solitons \citep{mikh85}. \cite{petrova04} proposed  generation by  induced scattering of waves leading  to a redistribution of the radio emission in frequency.   \cite{Weatherall,Hankins03} proposed  strong plasma turbulence; \cite{eilek07}   discuss a possibility of  interference fringes. In spirit, our work is closest to \cite{Istomin}, who proposed that GP are generated on  \Alfven waves  during a reconnection event close to the light cylinder. On the other hand, discovery of the banded structure of IGPs  is inconsistent with emission on  \Alfven waves (see Appendix \ref{other}), and, in addition,  \Alfven cannot leave \ms and need to be converted into escaping modes.

\begin{figure}[!h]
     \includegraphics[width=1.0\linewidth]{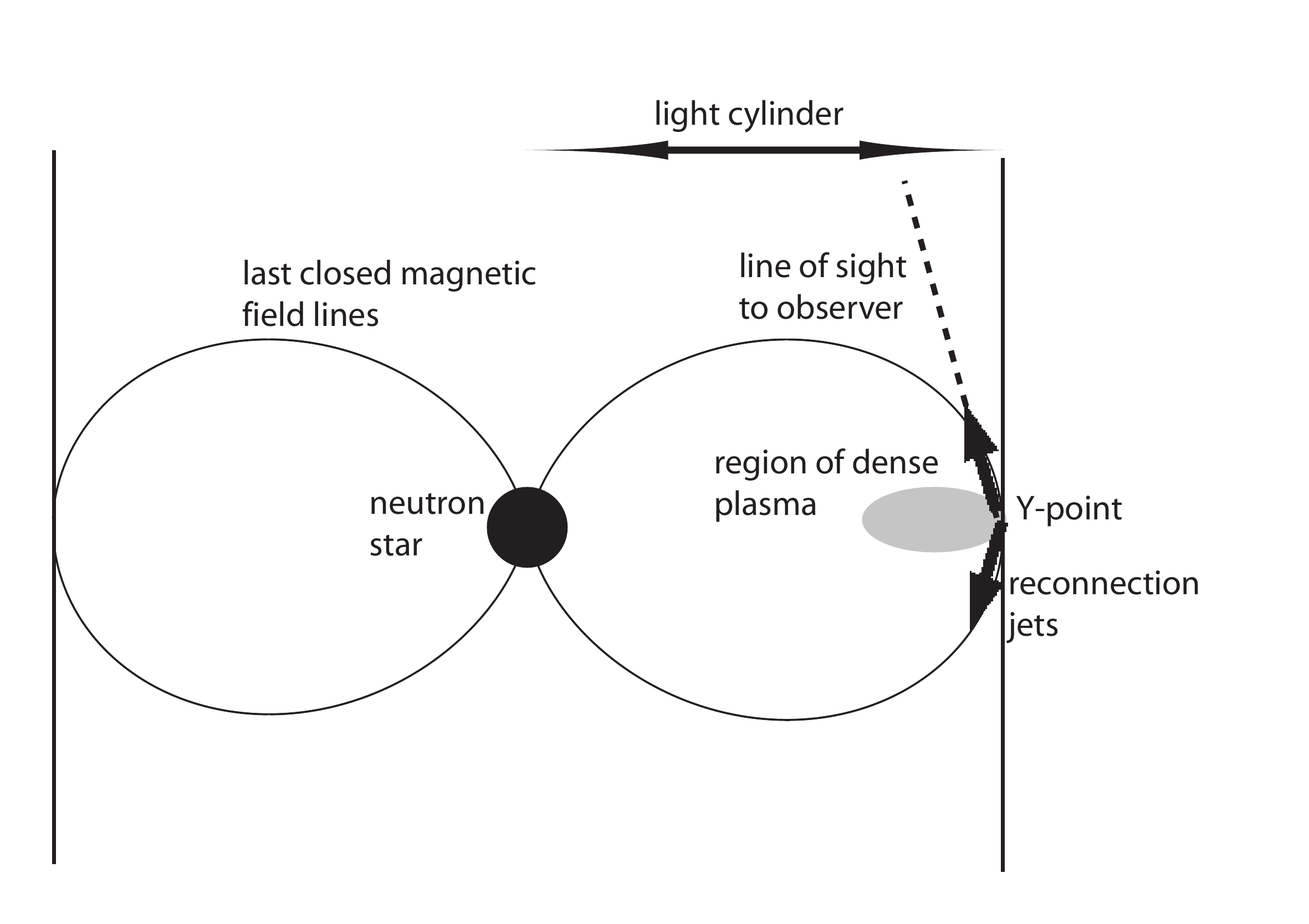}
   \caption{Generation region of giant pulses in Crab. High density plasma is trapped on closed field lines near the light cylinder. Occasional reconnection jets produce high Lorentz factor beams that  propagate along \Bf lines and  emit  coherent cyclotron-Cherenkov  radiation at  anomalous Doppler resonance.}
 \label{dipole}
 \end{figure}

Before laying out the details we give here a brief overview of the model, Fig. \ref{dipole}.
 The model requires that a high density plasma, with density $\sim 10^5$ times Goldrech-Julian density, is present on closed field lines. In addition, an occational high energy beam, with  radiation-limited Lorentz factor 
 $\sim 10^7$, is propagating along the field.  These requirement come from actual fitting of observed 
 emission bands to a particular cyclotron resonance conditions. The fit requires that the bulk plasma be not moving relativistically (hence association with closed field lines), the value of \Bf be the minimum 
 possible in the \ms (even the factor of two difference of \Bf strength  between the magnetic pole and equator are important), emission region be limited to small volume (hence association with the Y-point).

\section{Wave dispersion in pulsar magnetosphere}

To guide the reader through the following derivations, we first  re-derive the basic properties of  wave dispersion in  pair plasma  in a strong \Bf of pulsar \ms. 
In the  standard model of pulsar magnetospheres 
\citep{GJ} rotating, strongly magnetized
 neutron stars induce strong electric fields that pull
the charges from their surfaces. Inside the closed  field
lines of the neutron star magnetosphere, a steady charge
distribution is established,  compensating the induced electric
field. It is typically  assumed that on closed field lines, particle density is at a minimum and equals charge density, the 
 Goldreich-Julian density $n_{GJ}= \Omega B/( 2 \pi c e)$.
In fact, the real particle density may  and does exceed this minimum value, as we argue in this paper. 
There is already a clear evidence that in case of the 
 double pulsar PSR  J0737$-$3039 plasma density on closed field line exceeds the Goldreich-Julian density  by a factor
$\sim 10^4-10^5$  \citep{lt05}, though it was not clear if this is  specific to the double pulsar.  
 One of the implications of the present model is that high density plasma is present on closed field lines
of isolated pulsars as well.

As a simple parametrization, we normalize the plasma density to the Goldreich-Julian density, 
$n_p = \lambda\, n_{GJ} $.
(We neglect  the  thermal motion of plasma particles  and drifts resulting form curvature of field lines and inhomogenuity of \Bf).
The surface polar \Bf of Crab pulsar is $3.7 \times 10^{12}$ G. 
Below we will argue that  {\it  Crab GPs are generated  near a magnetic equator}, where \Bf at a given radius is two times smaller. Since the resonant frequency is proportional nearly to the third power of the local \Bf (\ref{X}), this factor of 2 is important.
Thus,  near the light cylinder  the cyclotron frequency at the equator is
\be
\om_B = { 9.5 \times 10^{12} \over
(r/R_{LC})^3 } {\rm rad s ^{-1}}
\label{omB}
\ee
For  the density parameter $\lambda= 3.3 \times  10^5$ (see below), the corresponding plasma frequency is 
  \be 
\om_p= \sqrt{ 2 \lambda \Omega \om_B} = {3.4 \times 10^{10} \over
(r/R_{LC})^{3/2} }{\rm rad s ^{-1}}
\label{omp}
\ee
($\nu_p =\om_p/(2 \pi) =5.5  \times 10^9 $ Hz).
Thus, near the light cylinder $\om_p/\om_B \sim 2 \times 10^{-3} $ (this is the maximum value of this parameter as a function of $r$.

\begin{figure}[h]
\includegraphics[width=0.95\linewidth]{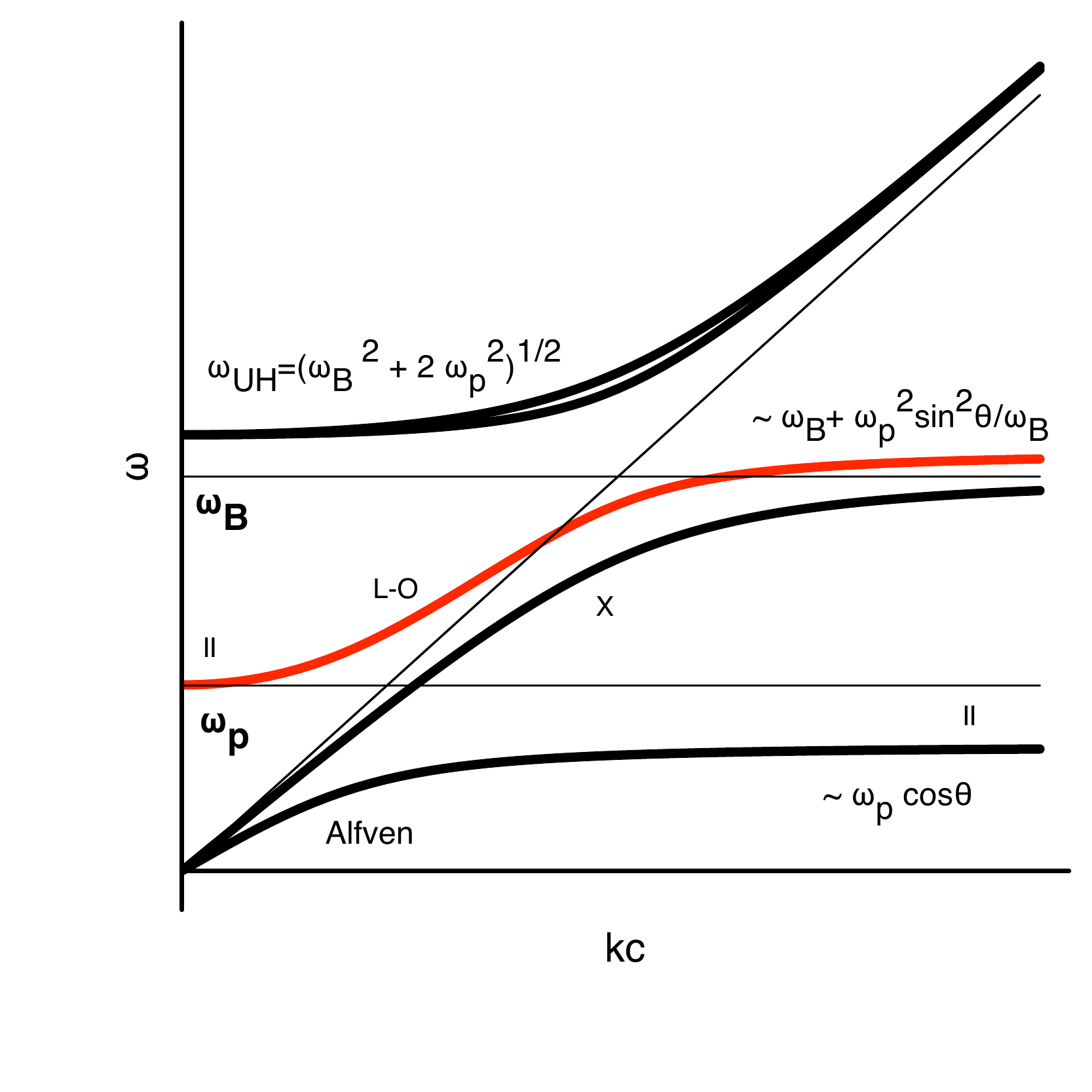}
\caption{Wave dispersions  $ \om(k) $ in pair plasma in strong \Bf, $\om_B \gg \om_p$, for oblique propagation. At low frequencies $\om \ll \om_B$ there are three modes labeled X (polarized orthogonally to   {\bf k} -{\bf B} plane), \Alfven and  L-O  (both polarized in the   {\bf k} -{\bf B} plane). The  L-O mode has a resonance at $\sim \om_B + \om_p^2 \sin^2 \theta/\om_B$ and cut-off at $\sqrt{2} \om_p$. 
The \Alfven mode has a resonance at  $\sim \sqrt{2} \om_p \cos \theta$. The sign $\parallel$ indicates locations where corresponding  waves are nearly longitudinally polarized. The two high frequency, $\om > \om_B$, waves with nearly identical dispersion
have a cut-off at the upper hybrid frequency $\om_{UH}=\sqrt{\om_B^2 +2 \om_p^2}$.
}
\label{1}
\end {figure}

For  $e^\pm$ plasma in \Bf the  dispersion relation factorizes giving two modes: the
X mode  with the
electric vector perpendicular to the {\bf k-B} plane 
and two branches of the
longitudinal-transverse mode, which we will call 
 L-O
and Alfv\'{e}n waves,  with
 the electric vector in the {\bf k-B} plane \citep[][see Fig. \ref{1}]{AronsBarnard86}.
 X waves is a
 {\it subluminal}  transverse electromagnetic wave with a dispersion relation
 \be
n^2 = 1 - { 2 \om_p^2 \over \om^2 -\om_B^2}
\ee
here $n= kc/\om$ is refractive index, $\om_B =e B/mc $ is cyclotron frequency,  $\om_p =\sqrt{ 4\pi n_{\pm} e^2/m}$ is a plasma frequency of each species (so that for pair plasma the total plasma frequency is $\sqrt{2} \om_p$).
The \Alfven-L-O mode satisfies the 
 dispersion relation
\be
n^2 = { (\om^2 -  2 \om_p^2)(\om^2 -  2 \om_p^2 - \om_B^2) 
\over (\om^2 -  2 \om_p^2) (\om^2 - \om_B^2 ) - 2 \om_B^2 \om_p^2 \sin^2 \theta }
\ee
 Alfv\'{e}n branch is always subluminal  while L-O mode
is {\it superluminal} at small wave vectors and 
{\it subluminal} at large wave vectors. 

\subsection{L-O mode}

As we argue that the GPs are generated on the L-O mode, let us discuss its properties in more detail. The
L-O mode exists in a frequency range
\be 
\sqrt{2} \om_p < \om < \left( {1\over 2} \left( \om_B^2 + 2 \om_p + \sqrt{ (\om_B^2 + 2 \om_p )^2 - 8 \om_B^2 \om_p^2 \cos^2 \theta} \right) \right)^{1/2}  \approx \om_B+{\om_p^2 \sin ^2 \theta \over \om_B}
\ee
It has a reflection point at $ \sqrt{2} \om_p$ and resonance at the upper bound.
L-O mode becomes luminal  at  frequency    $\om_0 = \sqrt{ 2 \om_p^2 +  \om_B^2 \sin^2 \theta}$.  Its polarization is 
\be
{e_x\over e_z} = { (\om^2 -\om_B^2) (\om^2 -2 \om_p^2) \cot \theta \over 
\om^2 (\om^2 -\om_B^2 - 2\om_p^2)}
\label{polariz}
\ee
(for wave vector in $x-z$ plane), 
so that 
the L-O mode is nearly longitudinal for $ kc \ll \om_0 $ (this becomes a Langmuir wave for parallel propagation) 
and is nearly transverse for $ kc \gg \om_0 $.  
At the luminal point $\om_0 $ polarization is quasi-longitudinal for
$\theta \ll (\om_p/\om_B)^2 $ and is quasi-transverse for larger angles. 

\section{Generation of giant pulses with banded frequency structure}

\subsection{Anomalous cyclotron resonance on L-O mode}
The
anomalous cyclotron resonance condition  is 
\be
\om- k_\parallel v_\parallel = s \om_B/\gamma, \mbox{ for $s < 0$}
\label{anom}
\ee 
where $k_\parallel$ and $ v_\parallel $ are components of wave vector and particle's velocity along \Bf,
  $\gamma$ is Lorentz factor of fast particles and
$s$ is an integer. 
It is clear that the necessary requirement for the anomalous cyclotron resonance
is that the refractive index of the mode be larger than unity, and that the parallel speed of the particle be larger than the wave's phase speed. The physics of emission is similar to the Cherenkov  process, except that during photon emission a particle {\it increases} its gyrational motion and goes {\it up} in Landau
levels \citep{Ginzburg}. The energy is supplied by parallel motion. Importance of anomalous cyclotron resonance for pulsar radio emission has been discussed by \cite{mu79,Kaz91,lbm99}. In these papers,  emission was argued to be generated on  open field lines at large distances from the start. 

The resonant  frequencies corresponding to  anomalous cyclotron resonance  for the X and L-O modes  for our fiducial parameters (see below) are shown in Fig. \ref{resfreq}. 
\begin{figure}[h]
\includegraphics[width=0.95\linewidth]{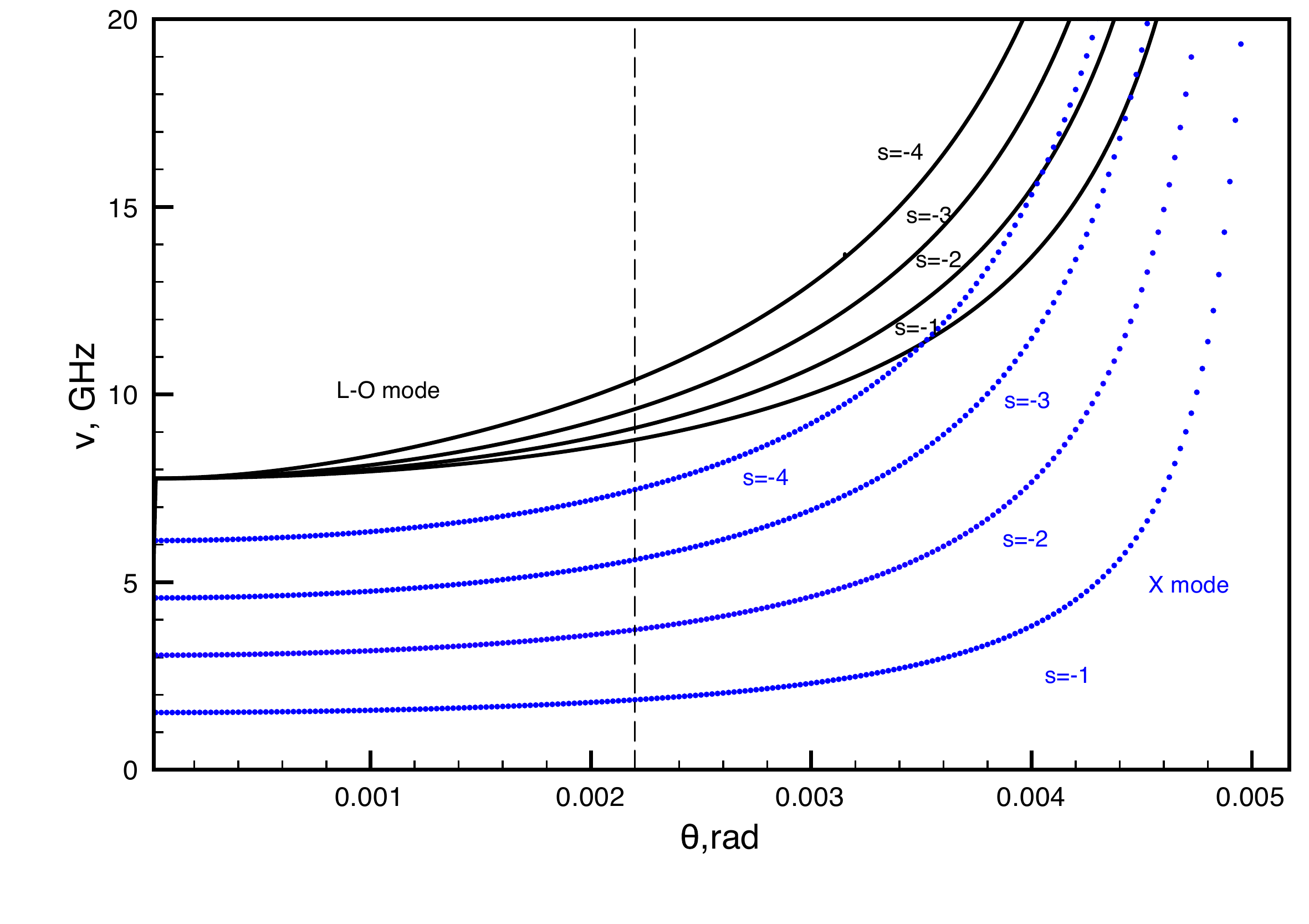}
\caption{Resonant frequencies for the L-O   (solid line)  and the X modes (doted line) for anomalous cyclotron resonance  as function of angle between the line of sight and \Bf for first four resonances $s=-1,... -4$. 
The cut-off  of the L-O mode at
$\theta =0$ is at total 
plasma frequency  $\sqrt{2} \om_p/(2 \pi) = 7.76 $ GHz. The X modes bands are equally spaced at $\theta=0$ with  $ \delta \nu = {\om_B^3 \over \gamma \om_p^2}$
(corresponding to $1.52$ GHz); higher order resonances are not shown. The vertical dashed line gives  the best fit values of $\theta = 0.0022$ for resonances on L-O mode. }
\label{resfreq}
\end {figure}
Overall, 
L-O mode dispersion is quite complicated, depending   sensitively on the small parameters $\om_p/\om_B$ and
$\theta$. As a guide we can use a much simpler condition for anomalous cyclotron resonance on X mode, keeping in mind
that 
X mode has phase velocities larger than L-O mode.
In the limit
$\om \ll \om_B$ resonance on X mode occurs at frequencies
\be
{\om} \sim {  |s|  \om_B^3 \over \gamma \om_p^2} 
\label{X}
\ee
Note, that  for $\gamma  \gg (\om_B/\om_p)^2$ both the  resonant frequency and frequency differences are   much smaller than cyclotron frequency.
 In case of relativistically streaming plasma  on open field lines,  the rhs of Eq. (\ref{X}) is multiplied by $\gamma_p^3$ (where $\gamma_p$ is Lorentz factor
 of plasma bulk motion),  as plasma density, wave frequency and Lorentz factor are all smaller by $\sim \gamma_p$. As a result, 
 for the fast, high field pulsar like Crab the resonance condition for radio waves cannot be satisfied inside the \ms for reasonable plasma parameters.  From (\ref{X}) it follows that in order to have
 resonant frequency below $\sim 10$ GHz, it is required that 
 $\gamma \lambda \geq 10^{13}$. This provides an order-of-magnitude estimate for the required plasma parameters. In making this estimate we neglected the Lorentz factor arising  due to rotational velocities of emitting plasma, which can be important near the light cylinder.

\begin{figure}[h]
\includegraphics[width=0.95\linewidth]{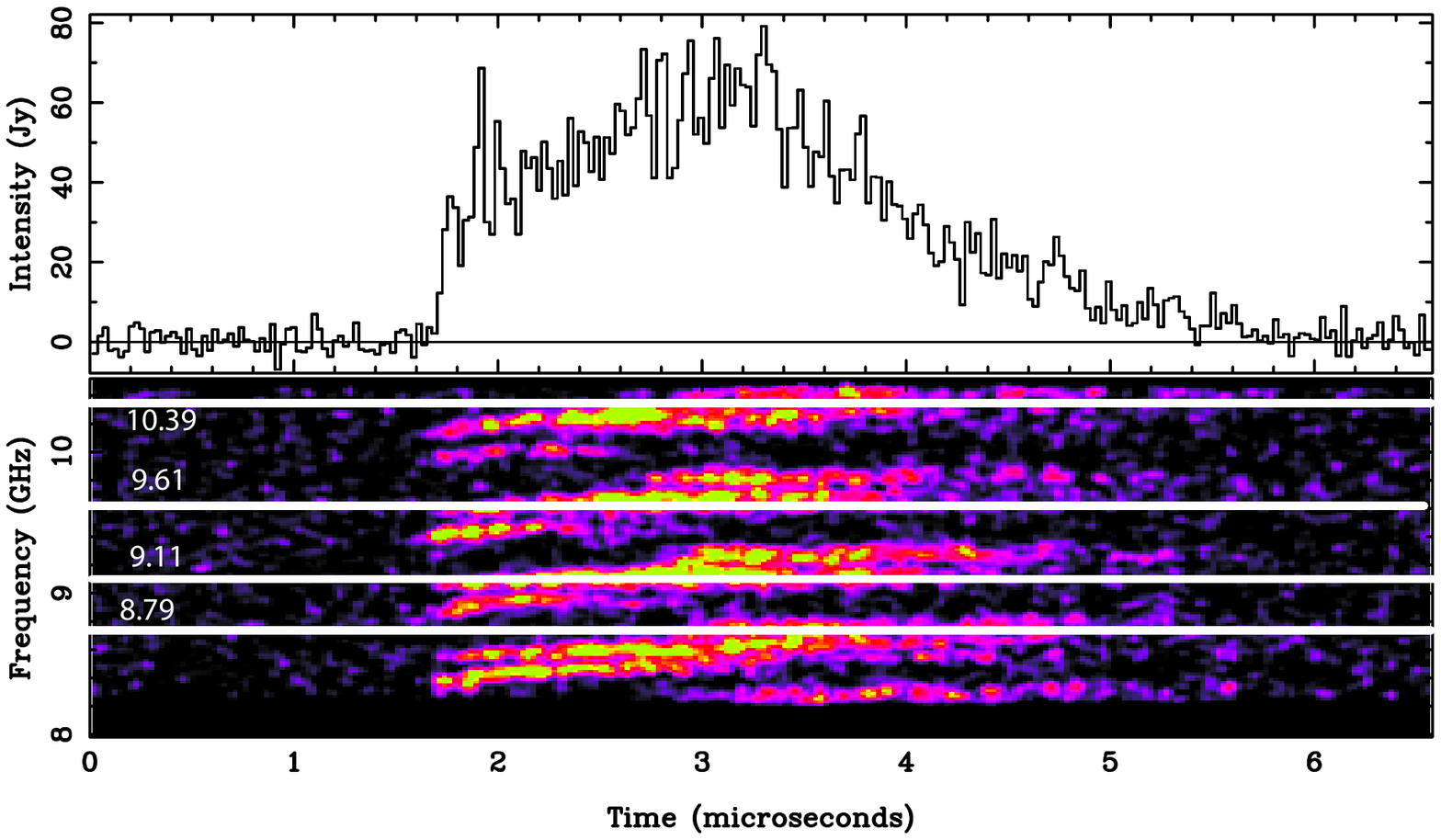}
\caption{Location of emission bands  (white stripes) for the fiducial model.  The fitted observations correspond to  Fig. 4 of \cite{eilek07}.
}
\label{Fit-Bands}
\end {figure}

Using the value of the \Bf at the light cylinder  (\ref{omB}) and parametrization of plasma density (\ref{omp})
we search through three parameters: plasma density normalization $\lambda$, Lorentz factor of the fast particles $\gamma$ and viewing angle $\theta$. 
To fit the results, we use Fig. 4 of \cite{eilek07}, trying to reproduce the frequencies and separations of the emission bands. 
We find a satisfying fit for the following parameters, see Fig. \ref{Fit-Bands}. 
Plasma over-density 
$\lambda = 3.3 \times 10^5$ (note, that this value is close to what was inferred for the plasma density on closed field lines
of pulsar B in the double pulsar PSR  J0737$-$3039  \citep{lt05}). Lorentz factor of the fast beam is
 $\gamma= 7.4 \times 10^7$; 
 this turns out to be close to the   estimate  of radiation-limited  acceleration (\ref{gamma}). 
 The viewing angle with respect to \Bf is
  $\theta =0.0022$.
  A reasonable fit is achieved in a fairly limited  volume of parameter space,   $  3.2 \times 10^5<\lambda <   3.8 \times 10^5$, $6.9 \times 10^7 < \gamma < 7.4 \times 10^7$,
$0.00251 < \theta < 0.00225$.
Most importantly, 
for the first four bands the value of bands' separation  $ 2 (\nu_{s+1}-\nu_s)/(\nu_{s+1}+\nu_s)$ are $.036, 0.053, 0.077$,
which is  very close to the mean quoted value of $0.06$.


\subsection{Growth rates}

Since for our best fit parameters $\theta \gg (\om_p/\om_B)^2$, Eq.  (\ref{polariz}) implies  the L-O mode is nearly transverse at resonance. We can then  use the previously calculated growth rates for excitation of transverse electromagnetic modes at anomalous cyclotron resonance
\citep{Kaz91,lbm99}
\be
\Gamma \sim {\om_{p,res}^2 \over \om}
\label{Gamma}
\ee
where $\om_{p,res}$ is the plasma frequency of the fast beam, generated during reconnection. 
Note, that since the emission  bands' spacing is much smaller than their carrier frequencies, the growth rates are nearly the same for different bands (observed intensity also depends on saturation mechanism). Similar growth rate and, presumably, similar saturation levels explain generation of multiple emission  bands.   We expect that during  a reconnection event, all the particles in the reconnection region may be accelerated, $\om_{p,res} \sim  \om \sim \om_p$, so that  the Growth rate can be very high 
$\Gamma \sim \om_p$

\subsection{Lorentz factor of fast particles  on closed lines}

We envision that reconnection events close to the Y-point (near the  last closed field lines at magnetic equator) accelerate  particles  along the closed field lines. 
The maximum available potential is related to the resistive process occurring during field reconnection, 
like the speed of reconnection and a number of flux tubes being reconnected. In addition, for nearly orthogonal rotators the  ''null charge surfaces" \cite{cr77}, where Goldreich-Julian density vanishes, lies inside the polar cap, so that  after  reconnection a  large potential drop would develop along the field lines \cite{Istomin}.
The upper bound,
which is not reached by far, can be estimated as  the total potential across the open field lines,
$\Phi \sim B_{NS} R_{NS} ( \Omega R /c)^2 $; this  corresponds to a Lorentz factor
 $\gamma_{max} \sim 10^{11}$ (\cite{Istomin} esimated  Lorentz factor is about an order of magnitude smaller). The maximum electron
energy will, in fact, be limited by radiative damping, like curvature and IC radiation.
Assuming that curvature radiation is the dominant loss process, we can estimate the maximum energy than an electron can reach in accelerating  electric field $E\sim  B$ (assuming that  reconnection inflow velocity is close to the speed of light). We find
\be
\gamma \sim\left({B\over e}\right)^{1/4} \sqrt{R_c} \sim 6.7 \times  10^7 \left({ r \over R_{LC}}\right)^{-1/4}
\label{gamma}
\ee
where $R_c \sim c/\Omega$ is the curvature radius. Surprisingly, this matches nearly exactly the value that we inferred from fitting the observed narrow-band structure.


\subsection{High plasma density on closed field lines}

According to the model, a high density plasma should be present on closed field lines. The overdensity we find here, modeling Crab GPs,  $\sim 10^5$, turns out to be close to the one found by eclipse modeling of the   binary pulsar system PSR J0737-3039A/B \cite{lt05}. How related are these peices of evidence and 
how  dense plasma  is created  and supported on closed field lines? These two cases   seem to very different, in fact. In slow pulsars, like PSR J0737-3039B, plasma can be efficiently trapped on closed field lines by magnetic bottling (for  $ \sim 10^6$ periods in case of   PSR J0737-3039B, \cite{lt05}), but not in case of Crab, in which case radiative decay times are too short, $\sim 10^{4} $ sec at the light cylinder. \cite{lt05}  proposed that high densities on closed field lines of PSR J0737-3039B may be explained by interaction with the wind of the companion, but similar overdensity in an isolated pulsar questions that possibility. Though there are several ways particles can populate closed field lines
 (\eg\ kinetic drift from open field lines, pair production by the 'backward" beam from outer gaps), the demands  in  case of short pariod pulsars like Crab are pretty high: the mechanism should create an over-density of the order $10^5$ with no efficient bottling.

 In case of Crab  this over-density is required in a fairly limited region near the light cyllinder, where IGPs are presumably produced. It is somewhat natural to associated this density enhancement with the
 Y-point, where the last closed field line approaches the  magnetic equator. In fact, we indeed may expect high density around that region. 
 First, even in rigidly rotating dipolar \ms\  the charge density diverges at that point \cite{GJ}. 
 Second,
   in case of a more realistic force-free aligned  rotator,  poloidal   \Bf\   also diverges at the Y-point \cite{Gruzinov,Spitkovsky06}. To resolve this divergency, non-ideal effects such as resistivity \cite{Spitkovsky06}  and/or particle inertia  \cite{Komissarov06} should be taken into account.  At the moment, we leave this possibility to further studies.

\subsection{Predictions: simultaneous  GLAST signal}

The model has interesting prediction for the forthcoming GLAST mission. The high energy beam is expected to produce curvature radiation at $ \sim  \hbar \gamma^3 \Omega \sim  30 $ GeV. This energy corresponds to the maximum effective area of the LAT instrument.  Thus, one expects to see GeV photons  contemporaneous with GPs. Estimating the beam density $\sim n_{GJ}$ and emitting volume as  $\sim 0.1 R_{LC}^3$, the total power in curvature photons is $\sim 2 \times 10^36$ erg/sec, or $4 \times 10^37 $ photons/sec. Assuming isotropic   emission (this gives a lower limit on the observed flux), the expected  flux at Earth  is  $\sim 10^{-7}$ photons cm$^{-2}$ s$^{-1}$ during a GP. This is nearly three orders of magnitude higher than the integral sensitivity of LAT at 30 GeV (for high latitude sources; Galactic background will also contribute in case of Crab), implying that a duty cycle of $\geq 10^{-3}$ for GP emission is needed to detect GP in GeV range.  We conclude that we may reasonably  expect  GeV photons contemporaneous with  GPs.

\section{Implications of the model}

Recent advances in observational technique uncovered  very  detailed properties of pulsar radio emission, and of Crab giant pulses in particular \citep{eilek07}. 
Inspired by  this work, in  the present  paper we describe a  quantitative model of Crab giant pulses. To the best of our knowledge this is the first work on pulsar emission mechanism that actually does a fitting of pulsar spectra to a particular model. The model can reproduce fairly well the  narrowband, unequally spaced emission bands  seen in GHz frequency range  in Crab giant pulses associated with the interpulse.
It also  explains  why banded structure in  IGPs in Crab  is  not seen below $ \sim 4$ GHz:
anomalous cyclotron resonance on L-O mode  occurs above plasma frequency, which for our typical parameters is in the GHz range. Pulse to pulse variation in location of emission bands are due to fluctuations of the plasma density.

Note, that though the emission involves cyclotron transitions, both the emission frequencies and inter-band spacing are  much smaller than the local cyclotron frequency for the best fit parameters. This is a curious property of anomalous cyclotron resonance.

Perhaps the most striking features of the model is that GPs are 
 generated on {\it closed} field lines, in stark contrast to all other models of pulsar radio emission. We stress that this only applies to GPs, regular pulses are generated on open field lines, as  is well established by a multitude of observational facts. We argue that giant pulses are different.
The fact that we see GPs in some particular pulsars  and that only the Crab IGP shows a banded frequency structure, is, in some sense, just a chance occurrence.
The resonance condition is a sensitive function of the angle between the line of sight,  local \Bf and plasma density, so that a mismatch by $\sim 10^{-3}$ radians in observer angle  either destroys the resonance completely or places the resonant frequency outside of the observed frequencies.  This explains why GPs are typically seen in narrow phase windows. This may  also be related to the fact that only few pulsars show GP phenomenon: in those cases we are lucky in terms of orientation, so that the line of sight passes close to the direction of local \Bf at the reconnection region, and in terms of local plasma parameters, so that the emission frequency, which scales nearly $\propto \om_B^3$ falls in the observed frequency range.
Exact location of the reconnection region is not specified in the model, but it is naturally to assume that it occurs close to the 
 Y point, where we expect  high current concentrations \citep{Gruzinov}.

One of  the main implications of the model is that (at least near the Y-point)
 closed field lines are populated with dense plasma, exceeding  by a factor $10^5-10^6$ the minimum 
charge density.  On the one hand, this is not expected from the basic pulsar model \citep{GJ}. On the other hand, presence of such dense plasma on closed field lines is solidly established in  the case 
of the double pulsar PSR  J0737$-$3039 \citep{lt05}. Somewhat surprisingly, the over-density $\sim 10^5$ (with respect to Goldreich-Julian density) we find here, modeling the generation of Crab giant pulsars, is similar to the over-density required to explain eclipses in the double pulsar.

To fit our findings into a global model of pulsar radio emission, we note that at frequencies above 2 GHz (but below NIR), Crab emission around the interpulse undergoes drastic changes with respect to both  lower radio frequencies and higher optical through X-ray emission \cite{hankins96}: the  interpulse is shifted by $\sim10 $ degrees and  new  High Frequency Components (HFC1 and HFC2)   trailing the interpulse appear. This is the range where the banded structure appears in IGPs. Our findings are consistent with the possibility that interpulse emission at these frequencies is a qualitatively different component from the main pulse and from the  low frequency interpulse. Still  a clear (geometrical?)  picture of emission components  is still missing.

 For small angles of propagation with respect to \Bfs, the dispersions of the X and the L-O modes nearly coincide, while polarization of both are nearly orthogonal for the parameters of interest. We have shown that the pattern of emission bands seen in IGPs 
 matches well the anomalous cyclotron resonance on the
  L-O mode. On the other hand,  resonance bands of the X mode  may lie fairly close to the L-O modes, so it is natural to expect that the
X mode should also be emitted.
Emission bands associated with the X modes have distinctly different properties from 
L-O mode, being spaced in nearly equal intervals, equal to the frequency of the first harmonic.
Large  
 spacing,  nearly the same as carrier frequency,  makes it nearly impossible to observe X-mode  emission bands with a finite receiver bandwidth (additional complications at low frequncies come from interstellar scattering).  Still, the fact that a spectrum is not continuous, but consisting of emission bands, may be verified statistically during simultaneous observations of GPs in different frequency bands. Such observations 
of the millisecond pulsar B1937+21 \citep{popov} showed that out of 10-15 GPs observed at two frequency windows 1414Ð1446 MHz and 2210Ð2250 MHz, no events were found to occur simultaneously at both frequencies. This requires that spectra of individual GPs are narrow band, $\Delta \nu/\nu \leq 0.5$,
consistent with our picture.

Additional complications may come from absorption of the X-mode at cyclotron resonance.
 Since at the light cylinder the radiation frequency is much smaller than the local cyclotron frequency,  resonant absorption may occur only in the wind \citep{Sincell}. At distances much larger than the light cylinder,  the  radially propagating L-O mode is polarized  nearly 
{\it along } \Bf and thus is not absorbed at the  cyclotron resonance \citep{Petrova06b}, while the X mode may experience  absorption.

The present  model can also accommodate qualitatively the fact that sometimes emission bands seem to be drifting slightly  up in frequency. If the fast beam is generated close to the magnetic equator, where \Bf is the weakest, then as it propagates along the closed field  lines, the local cyclotron frequency increases, leading to an increase of the observed frequency. Thus, placing beam generation region at magnetic equator explains why emission bands always drift up in frequency (Eilek \& Hankins, priv. comm.).

At the moment it is not clear how rare or ubiquitous is the phenomenon of GPs in pulsars. As many as tens percent may show GPs (Ransom, priv. comm.). Of those pulsars that do show GPs,
detection of banded structure in GPs depends sensitively on the plasma parameters close to the light cylinder and on the 
angle between line of sight and local \Bf.   Taking our best fit value of $\theta \sim 2 \times 10^{-3}$ we can estimate a relative number of pulsars  in which we expect to see bands in  GPs as  $\sim \theta$.
 Such an  estimate is very rough and  is naturally subject to a number of uncertainties, like beaming fraction of normal and giant pulses.

Finally,  we note that closed field lines of the Earth \ms are very active in producing high brightness radio emission
like auroral hiss,  roars and burst \citep{Labell02}, though they all occur at normal cyclotron resonance 
(and thus require a  loss cone-type distribution) at frequencies of the order of the local cyclotron frequency 
(and thus are too high to be directly  applicable to observations of Crab pulsar at radio waves). It is intriguing to speculate that similar type  emission might be observed from Crab at $\sim 10^{12} -10^{13}$ Hz, but  the required observations at such frequencies are challenging.

I would like to thank Timothy Hankins, Jean Eilek, Qinghuan Luo,  George Machabeli, Alison Mansheim, Mikhail Popov  and Scott Ransom for discussions that were vital for this work.

\bibliographystyle{apj}
\bibliography{PulsarBib}

\appendix
\section{Other excitation mechanisms}
\label{other}

In this appendix we discuss other excitation mechanisms and point out why they cannot be responsible for Crab GPs, at least for those  GPs that show a narrowband structure. In doing so, we assume that the observed band structure is due to emission processes, not propagation.

At $\om \ll \om_B$ there are three modes: \Alfven, X and L-O, which can be excited at Cherenkov (except  X mode), cyclotron and anomalous cyclotron resonances.
We have discussed excitation of the X and the L-O modes at  anomalous cyclotron resonances; let us now discuss other possibilities.

\subsection{Normal cyclotron resonance}

For instability to operate on normal cyclotron resonance particle distribution should be of the loss-cone type. 
Such distributions may be  expected in the outer gap models, which produce secondary particles with non-zero transverse momenta. Effects of magnetic bottling on a down-ward propagating particles will  create a loss-cone. 
In addition, since an outer gap operates near the last closed field lines, it is expected that some particles will drift from open to closed lines. This possibility is attractive since the closed field lines may store particles bouncing between the poles for a long time. 
Development of the loss-cone instability
would then convert this stored energy into radiation. This would create radio emission from closed field lines. 
Both the X and  the L-O modes can be emitted at cyclotron resonance, but 
the main  problem is that emitted frequencies are typically close to cyclotron frequency $\om_B$, which is some three orders of magnitude higher than observed. In principle, the resonant  frequency 
can be smaller than $\om_B$ as, \eg, in case of  backward wave oscillators, but this would require fine tuning the frequencies in one part in a thousand. In addition, inter-band frequency spacing is large, 
$\sim \om_B$.

\subsection{\Alfven mode}

Excitation of \Alfven waves in pulsars has been considered by \cite{lyutAlf}.
\Alfven waves exist for $0< \om  <\left( {1\over 2} \left(  \om_B^2 + 2 \om_p - \sqrt{ (\om_B^2 + 2 \om_p )^2 - 8 \om_B^2 \om_p^2 \cos^2 \theta}\right)  \right)^{1/2} \approx  \sqrt{2} \om_p \cos\theta $.
At low frequencies, $\om \ll \om_p \cos \theta$, $\om_p \ll \om_B$,   dispersion relation  becomes 
\be
\om = kc \cos \theta \left( 1 - {\om_p^2 \over \om_B^2} - { (kc)^2 \sin^2 \theta \over \om_p^2} \right),
\mbox{ for $\om \ll \om_p \cos \theta$, $\om_p \ll \om_B$}
\ee

\Alfven waves can be excited at 
Cherenkov resonance at 
\be
k_{res} c = {\sqrt{2} \om_p \over \sin \theta} \sqrt{ {1\over \gamma_b^2} - { 2 \om_p^2 \over \om_B^2}}
\ee
which requires  $ \om_B /( \sqrt{2} \om_p \gamma_b) > 1$. The resonant frequency is
\be
\om \approx { \sqrt{2} \om_p  \over \gamma_b} \cot \theta,
\ee
so that 
at a given angle emission is generated with given $\om$.
Though the resonance condition can, in general, be satisfied, it not   clear how to produce different emission bands.

\begin{figure}[h]
\includegraphics[width=0.95\linewidth]{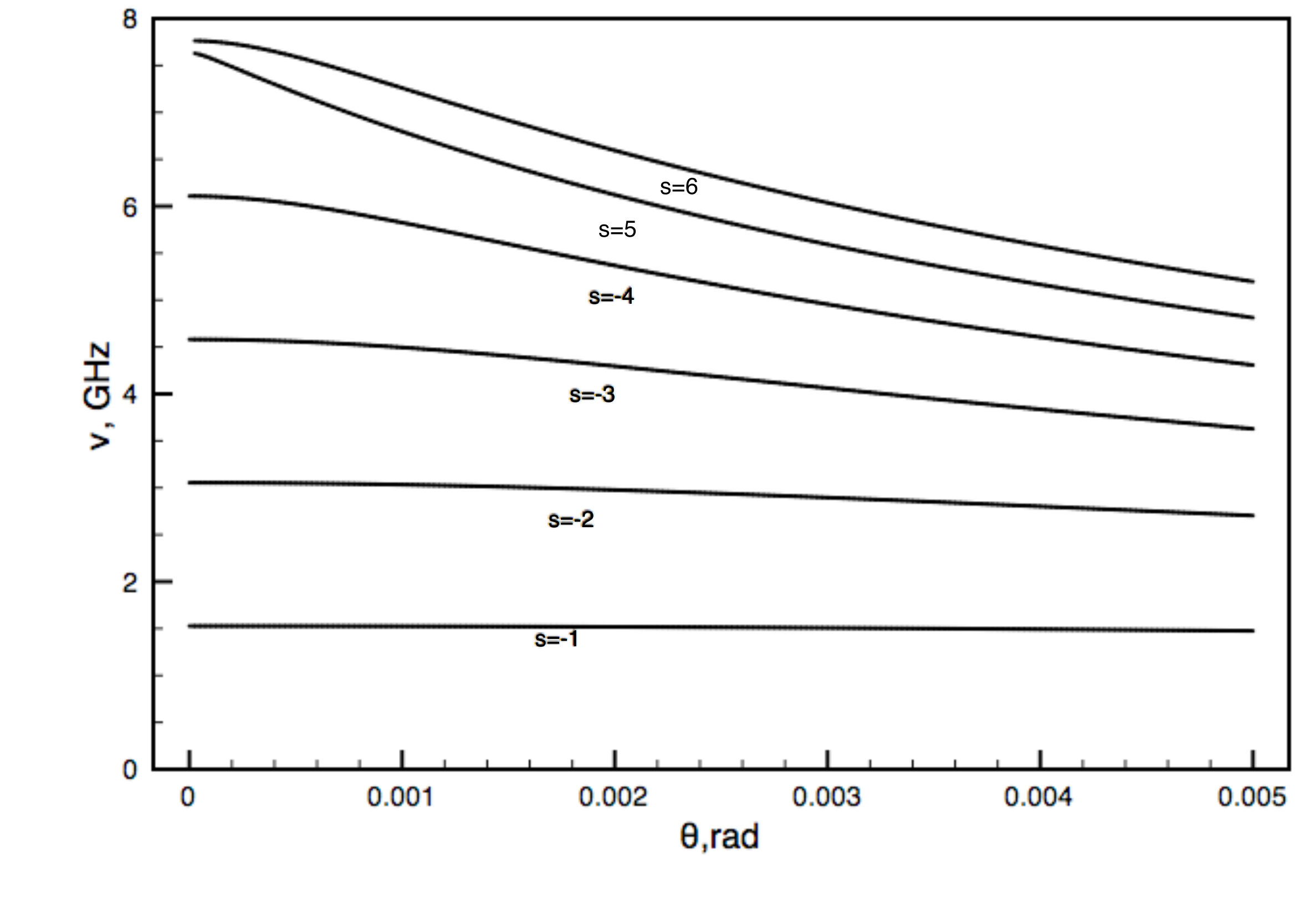}
\caption{Resonant frequencies for anomalous excitation of \Alfven waves. Parameters are the same as in Fig.  \ref{Fit-Bands}. \Alfven waves frequencies are limited to $\om < \sqrt{2} \om_p \cos \theta$.  Contrary to observations,  the inter-band spacing decreases with frequency.} 
\label{Alfv1}
\end {figure}

\Alfven waves can also be emitted at anomalous Doppler resonance. For small angles, $\theta \ll {\om_p^2 \over \om_B kc}$, the resonant frequencies are similar to those of X mode, while
 for $\theta \gg {\om_p^2 \over \om_B kc}$ we find
\be
k_{res} c =\left( {s \om_B \om_p^2\over \gamma_b \cos\theta \sin^2 \theta}\right)^{1/3}
\label{Alfv}
\ee
In addition, 
condition $\om \leq \om_p$ requires
\be
\theta \geq \sqrt{ \om_B \over \gamma_b \om_p}
\ee
 An appealing property of  the resonant condition(\ref{Alfv}) is that it produces a set of 
 frequency bands, separated by frequencies much smaller than cyclotron frequency $ \om_B$.
On the other hand, 
inter-band  separation {\bf decreases} with  frequency, see Fig \ref{Alfv1}, contrary to observations. 
This invalidates the model of  \cite{Istomin}, at least in application to IGPs.
In addition,  \Alfven waves cannot escape from plasma and need to be converted into escaping modes before they are damped on plasma particles.
 
\end{document}